\documentstyle[12pt,epsfig,axodraw,a4]{article} 
\def\bild#1#2{    
        \vspace*{-5mm}
        \begin{center}
        \begin{math}
        \epsfxsize#2cm
        \epsffile{#1}
        \end{math}
        \end{center}  }
\newcommand{\vs}{\vspace{-0.25cm}}
\begin{document} 
\begin{center}
\large{\bf Three-body spin-orbit forces from chiral two-pion exchange}

\bigskip

N. Kaiser\\

\medskip

{\small Physik Department T39, Technische Universit\"{a}t M\"{u}nchen,
    D-85747 Garching, Germany}

\end{center}

\medskip

\begin{abstract}
Using chiral perturbation theory, we calculate the density-dependent spin-orbit
coupling generated by the two-pion exchange three-nucleon interaction involving
virtual $\Delta$-isobar excitation. From the corresponding three-loop Hartree 
and Fock diagrams we obtain an isoscalar spin-orbit strength $F_{\rm so}(k_f)$ 
which amounts at nuclear matter saturation density to about half of the 
empirical value of $90\,$MeVfm$^5$. The associated isovector spin-orbit 
strength $G_{\rm so}(k_f)$ comes out about a factor of 20 smaller. 
Interestingly, this three-body spin-orbit coupling is not a relativistic effect
but independent of the nucleon mass $M$. Furthermore, we calculate the
three-body spin-orbit coupling generated by two-pion exchange on the 
basis of the most general chiral $\pi\pi NN$-contact interaction. We find 
similar (numerical) results for the isoscalar and isovector spin-orbit 
strengths $F_{\rm so}(k_f)$ and $G_{\rm so}(k_f)$ with a strong dominance of 
the p-wave part of the $\pi\pi NN$-contact interaction and the Hartree 
contribution. 
\end{abstract}

\bigskip

PACS: 12.38.Bx, 21.30.Fe, 24.10.Cn, 31.15.Ew\\


\bigskip 
The microscopic understanding the dynamical origin of the strong nuclear 
spin-orbit force is still one of the key problems in nuclear physics. The 
analogy with the spin-orbit interaction in atomic physics gave the hint that it
could be a relativistic effect. This idea has lead to the construction of the 
(scalar-vector) mean-field models for nuclear structure calculations 
\cite{walecka,ringreview}. In these models the nucleus is described as a 
collection of independent Dirac quasi-particles moving in self-consistently 
generated scalar and vector mean-fields. The footprints of relativity become 
visible through the large nuclear spin-orbit coupling which emerges in that
framework naturally from the interplay of the two strong and counteracting
(scalar and vector) mean-fields. The corresponding many-body calculations are
usually carried out in the Hartree approximation, ignoring the negative-energy
Dirac-sea. The NN-interaction underlying these models is to be considered as 
an effective one that is tailored to properties of finite nuclei but not 
constrained (completely) by the observables of free NN-scattering.  

On the other hand it has long been known that calculations based on 
Hamiltonians which contain only realistic two-nucleon potentials (thus fitting 
accurately all NN-phase shifts and mixing angles below the NN$\pi$-threshold) 
often cannot predict the observed spin-orbit splittings of nuclear levels. In 
fact one of the original motivations for the Fujita-Miyazawa three-nucleon 
potential \cite{fujita} was just the study of such spin-orbit splittings. In 
ref.\cite{pand} it has then been shown that one out of the Urbana family of 
three-nucleon forces makes a substantial contribution to the spin-orbit 
splitting in the nucleus $^{15}$N. Moreover, three-nucleon forces are
actually needed in addition to realistic two-nucleon potentials in order to
reproduce the correct saturation point of (isospin-symmetric) nuclear matter
\cite{akmal}. The long-range part of the three-nucleon interaction is generated
in a natural way by two-pion exchange \cite{pieper} and it can in fact be
predicted by using chiral symmetry \cite{friar}.  

The purpose of this paper is present analytical results for the nuclear 
spin-orbit coupling generated by the (chiral) two-pion exchange three-nucleon 
interaction. In order to arrive at such results we will make use of the 
density-matrix expansion of Negele and Vautherin \cite{negele}. This technique 
allows one to compute diagrammatically the nuclear energy density functional 
which includes the wanted spin-orbit coupling term proportional to the density 
gradient. We will first consider the three-nucleon interaction proposed 
originally by Fujita and Miyazawa \cite{fujita} where two pions are exchanged 
between nucleons while the third nucleon is excited to a (p-wave) $\Delta
$-resonance. We are able to evaluate the corresponding three-loop Hartree and 
Fock diagrams in closed analytical form. Then, we will turn to the two-pion 
exchange three-nucleon interaction generated by the (most general) chiral 
$\pi\pi NN$-contact vertex proportional to the second-order low-energy 
constants $c_j$. The effects from explicit $\Delta$-excitation reappear
in this description via resonance contributions to the low-energy constants 
$c_{3,4}$. In both approaches we will separately discuss the isoscalar and 
isovector spin-orbit strengths. 

Let us begin with writing down the explicit form of the spin-orbit coupling 
term in the nuclear energy density functional:
\begin{equation} {\cal E}_{\rm so}[\rho_p,\rho_n,\vec J_p,\vec J_n] = \vec
\nabla  \rho\cdot\vec J\, F_{\rm so}(k_f)+ \vec \nabla  \rho_v \cdot\vec J_v\, 
G_{\rm so}(k_f)\,, \end{equation} 
where the sums $\rho=\rho_p+\rho_n$, $\vec J= \vec J_p+\vec J_n$ and
differences  $\rho_v=\rho_p-\rho_n$, $\vec J_v= \vec J_p-\vec J_n$ of proton
and neutron quantities have been introduced.  
\begin{equation} \rho_{p,n}(\vec r\,) = {k_{p,n}^3(\vec r\,) \over 3\pi^2} = 
\sum_{\alpha \in \rm occ} \Psi^{(\alpha)
\dagger}_{p,n}( \vec r\,)\Psi^{(\alpha)}_{p,n}( \vec r\,)\,,\end{equation} 
denote the local proton and neutron densities which we have rewritten in terms
of the corresponding (local) proton and neutron Fermi-momenta $k_{p,n}(\vec
r\,)$ and expressed as sums over the occupied single-particle orbitals
$\Psi^{(\alpha)}_{p,n}( \vec r\,)$. The spin-orbit densities of the protons and
neutrons are defined similarly: 
\begin{equation} \vec J_{p,n}(\vec r\,)=\sum_{\alpha \in \rm occ}\Psi^{(\alpha)
\dagger}_{p,n}(\vec r\,)i\, \vec \sigma \times \vec \nabla\Psi^{(\alpha)}_{p,n
}( \vec r\,) \,. \end{equation} 
Furthermore, $F_{\rm so}(k_f)$ and $G_{\rm so}(k_f)$ in eq.(1) denote the 
density dependent isoscalar and isovector spin-orbit strength functions. In
Skyrme parameterizations \cite{reinhard} these are just constants, $F_{\rm so}
(k_f)=3G_{\rm so}(k_f)=3W_0/4$, whereas in our calculation their explicit
density dependence originates from the finite range character of the two-pion 
exchange three-nucleon interaction.  

The starting point for the construction of an explicit nuclear energy density 
functional ${\cal E}_{\rm so}[\dots]$ is the bilocal density-matrix as given by
a sum over the occupied energy eigenfunctions: $\sum_{\alpha\in \rm occ}\Psi^{(
\alpha)}_{p,n}( \vec r -\vec a/2)\Psi^{(\alpha) \dagger}_{p,n}(\vec r +\vec a/2
)$. According to Negele and Vautherin \cite{negele} it can be expanded in 
relative and center-of-mass coordinates, $\vec a$  and $\vec r$, with expansion
coefficients determined by purely local quantities (nucleon density, kinetic 
energy density and spin-orbit density). As outlined in Sec.\,2 of 
ref.\cite{efun} the Fourier-transform of the (so expanded) density-matrix 
defines in momentum-space a medium-insertion $\Gamma(\vec p,\vec q\,)$ for the 
inhomogeneous many-nucleon system. It is straightforward to generalize this 
construction to the isospin-asymmetric situation of different proton and 
neutron local densities $\rho_{p,n}(\vec r\,)$ and $\vec J_{p,n}(\vec r\,)$. 
We display here only that part of the medium-insertion $\Gamma(\vec p,\vec q\,
)$ which is actually relevant for the diagrammatic calculation of the isoscalar
and isovector spin-orbit terms defined in eq.(1): 
\begin{eqnarray} \Gamma(\vec p,\vec q\,)& =& \int d^3 r \, e^{-i \vec q \cdot
\vec r}\,\bigg\{ {1+\tau_3 \over 2}\,\theta(k_p-|\vec p\,|) +{1-\tau_3 \over 2}
\,\theta(k_n-|\vec p\,|) \nonumber \\ && +{\pi^2 \over 4k_f^4}
\Big[\delta(k_f-|\vec p\,|) -k_f \,\delta'(k_f-|\vec p\,|) \Big]\, (\vec \sigma
\times \vec p\,) \cdot(\vec J+\tau_3\,\vec J_v) \bigg\}\,.  \end{eqnarray}
When working to quadratic order in deviations from isospin symmetry
(i.e. proton-neutron differences) it is sufficient to use an average
Fermi-momentum  $k_f$ in the prefactor of the spin-orbit density $\vec J
+\tau_3\, \vec J_v$. The double-dash in the left picture of Fig.\,1 symbolizes
the medium insertion $\Gamma(\vec p,\vec q\,)$ together with the assignment of
the out- and in-going nucleon momenta $\vec p \pm \vec q/2$. The momentum 
transfer $\vec q$ is provided by the Fourier-components of the inhomogeneous 
(matter) distributions $\rho_{p,n}(\vec r\,)$ and $\vec J_{p,n}(\vec r\,)$.

\bigskip

\bild{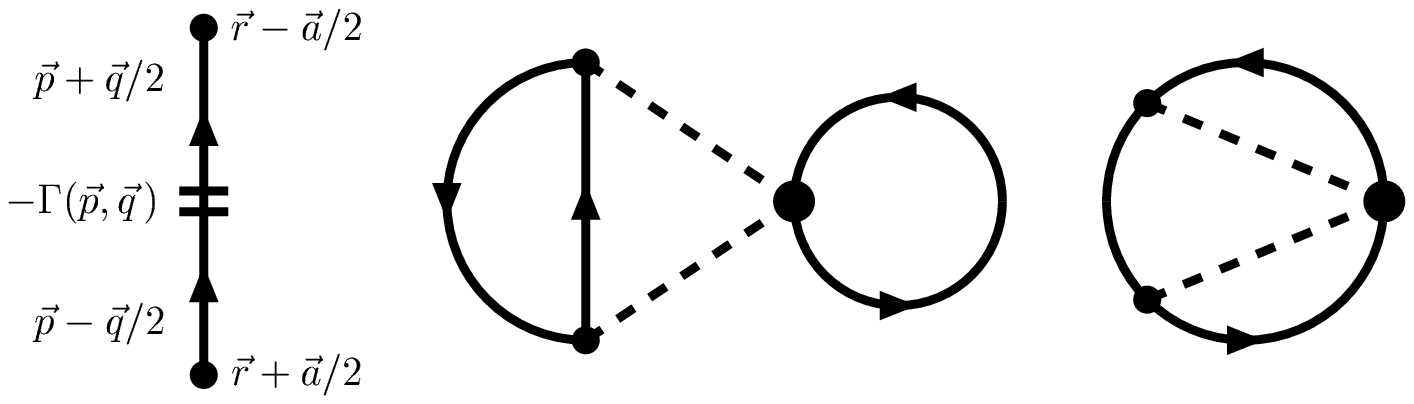}{12}
{\it Fig.\,1: Left: The double-dash symbolizes the medium insertion
$\Gamma(\vec p,\vec q\,)$ defined by eq.(4). Next shown are the three-loop 
two-pion exchange Hartree and Fock diagrams involving one chiral $\pi \pi
NN$-contact vertex (symbolized by the heavy dot). The combinatoric factors of
these diagrams are 1/2 and 1, in the order shown.} 

\bigskip

\bigskip

\bild{3bodyfig2.epsi}{16}
{\it Fig.\,2: Two-pion exchange Hartree and Fock diagrams with (single) virtual
$\Delta$-isobar excitation. The solid double-line denotes the $\Delta$-isobar 
and dashed and solid lines represent pions and nucleons, respectively. For
isospin-symmetric nuclear matter the isospin factors of these diagrams are 8,
0, and 8. The combinatoric factor is 1 in each case.}  

\bigskip

Now we turn to the analytical evaluation of the two-pion exchange diagrams with
(single) $\Delta$-isobar excitations shown in Fig.\,2. We give for each diagram
only the final result for the spin-orbit strengths $F_{\rm so}(k_f)$
and  $G_{\rm so}(k_f)$ omitting all technical details related to extensive
algebraic manipulations and solving elementary integrals. Putting a medium
insertion at each of the three nucleon propagators of the Hartree diagram (left
diagram in Fig.\,2) we obtain the following contribution to the isoscalar
spin-orbit strength:
\begin{equation} F_{\rm so}(k_f)^{(\Delta-\rm Hart)}= {g_A^4 m_\pi \over 8\pi^2
\Delta f_\pi^4} \bigg\{ {u+2u^3 \over 1+4u^2}-{1\over 4u}\ln(1+4u^2) \bigg\} 
\,, \end{equation} 
where $u=k_f/m_\pi$ denotes the ratio of the two small scales $k_f$ and
$m_\pi$. The $\Delta$-propagator shows up in this expression merely via 
the (reciprocal) $\Delta N$-mass splitting, $\Delta = 293\,$MeV. Additional
corrections to the $\Delta$-propagator coming from differences of nucleon 
kinetic energies etc. will make a contribution to the spin-orbit strength
$F_{\rm so}(k_f)$ at least one order higher in the small momentum expansion. 
In eq.(5) we have inserted the empirically well-satisfied relation $g_{\pi N
\Delta}=3g_{\pi N}/\sqrt{2}$ for the $\pi N\Delta$-coupling constant together 
with the Goldberger-Treiman relation $g_{\pi N} = g_A M/f_\pi$. Let us
briefly sketch the main mechanism which generates the strength function $F_{so}
(k_f)$. The exchanged pion-pair in the Hartree digram transfers a momentum 
$\vec q$ between the left and the right nucleon ring and this momentum $\vec q$
enters also the pseudovector $\pi NN$-interaction vertices. The spin-orbit 
strength $F_{so}(k_f)$ arises from the spin-trace tr$[\vec \sigma \cdot (\vec 
Q+ \vec q/2)\, \vec\sigma \cdot (\vec Q -\vec q/2)\,\vec \sigma \cdot(\vec
p\times \vec J\,)] = 2i(\vec q \times \vec Q\,)\cdot(\vec p \times \vec J\,)$
where $i\vec q$ gets converted to $\vec \nabla k_f =(\pi^2/2k_f^2)\vec \nabla
\rho$ by Fourier transformation. The rest is a solvable integral over the
product of three Fermi spheres. The second Fock diagram in Fig.\,2 (with
parallel pion lines) has the isospin factor $0$ for isospin-symmetric nuclear
matter and from the third Fock diagram (with crossed pion lines) we get the 
following contribution to the isoscalar spin-orbit strength: 
\begin{equation} F_{\rm so}(k_f)^{(\Delta-\rm Fock)}= {g_A^4 m_\pi u^{-3}\over
\pi^2 \Delta (16f_\pi)^4} \Big[8u^2-12 +(3u^{-2}+ 4)\ln(1+4u^2) \Big]^2 \,.
\end{equation}
It is highly remarkable that the pertinent nine-dimensional integral over the
product of two (different) pion-propagators and other momentum dependent 
factors can solved in terms of (a square of) elementary functions without the
occurrence of any dilogarithm. The specific isospin structures of the $\pi NN$-
and $\pi N\Delta$-vertices determine uniquely the ratio of isovector to 
isoscalar spin-orbit strength of each of the three diagrams. We find that the 
(left) Hartree diagram in Fig.\,2 does not contribute to the isovector 
spin-orbit strength $G_{\rm so}(k_f)$ while the combined result of both Fock 
diagrams in Fig.\,2 reads:  
\begin{equation} G_{\rm so}(k_f)^{(\Delta-\rm Fock)}= {7\over 3}\, 
F_{\rm so}(k_f)^{(\Delta-\rm Fock)} \,, \end{equation}
with a contribution of the second and third Fock diagram in the ratio $6:1$. It
is important to note that the expressions in eqs.(5,6) are independent of the 
nucleon mass $M$ and therefore these $2\pi$-exchange three-body spin-orbit
couplings are $not$ relativistic effects. In fact the diagrams in Fig.\,2 with 
two medium insertions (on non-neighboring nucleon propagators) do also generate
two-body spin-orbit couplings. The latter are however genuine relativistic 
effects proportional to $1/M$ and therefore counted as one order higher in the
small momentum expansion. Such two-body contributions to the spin-orbit
strengths $F_{\rm so}(k_f)$ and $G_{\rm so}(k_f)$ can generally be
expressed in terms of the spin-orbit amplitudes entering the T-matrix of
elastic NN-scattering: 
\begin{eqnarray} F_{\rm so}(k_f)^{(\rm 2-body)} &=& -{1\over 6} \bigg\{ 3
V_{SO}(0)+ V_{SO}(2k_f)+3 W_{SO}(2k_f) \nonumber \\ && + \int_0^1 dx\,x \Big[ 
V_{SO}(2xk_f) + 3W_{SO}(2x k_f) \Big] \bigg\} \,, \end{eqnarray} 
\begin{eqnarray} G_{\rm so}(k_f)^{(\rm 2-body)} &=& {1\over 6} \bigg\{  
W_{SO}(2k_f)-V_{SO}(2k_f)-3W_{SO}(0) \nonumber \\ && + \int_0^1 dx\,x \Big[ 
W_{SO}(2xk_f) -V_{SO}(2x k_f) \Big] \bigg\} \,. \end{eqnarray} 
The terms $-V_{SO}(0)/2$ and $-W_{SO}(0)/2$ belong to Hartree-type diagrams 
(with two closed nucleon lines) while the remaining ones summarize the
contributions from Fock-type diagrams (having just one closed nucleon line). 
Explicit expressions for the isoscalar and isovector spin-orbit NN-amplitudes 
$V_{SO}(q)$ and $W_{SO}(q)$ as they arise from $2\pi$-exchange with (single 
and double) $\Delta$-excitation can be found in the appendix of 
ref.\cite{delta} (modulo regularization dependent additive constants).  

For the numerical evaluation of eqs.(5,6,7) we use the (physical) parameters:
$M= 939\,$MeV (nucleon mass), $m_\pi = 135\,$MeV (neutral pion mass), $f_\pi =
92.4\, $MeV (pion decay constant) and $g_A = 1.3$ (equivalent to a $\pi 
NN$-coupling constant of $g_{\pi N} = g_A M/f_\pi = 13.2)$. The full line in 
Fig.\,3 shows the isoscalar spin-orbit strength $F_{\rm so}(k_f)$ generated by
the two-pion exchange three-nucleon interaction involving virtual $\Delta
$-excitation as a function of the nucleon density $\rho = 2k_f^3/3\pi^2$. 
As it is typical for a three-body effect the spin-orbit strength $F_{\rm
so}(k_f)$ starts from the value zero at zero density $\rho = 0$. The
contribution of the Hartree diagram is by far the dominant one. At nuclear 
matter saturation density (where $k_f \simeq 2m_\pi$) one finds for example 
$F_{\rm so}(2m_\pi)^{(\Delta-\rm Hart)} = 48.2\,$MeVfm$^5$ to be compared with
a Fock contribution of $F_{\rm so}(2m_\pi)^{(\Delta-\rm Fock)}=1.2\,$MeVfm$^5$.
Clearly, this $2\pi$-exchange three-body spin-orbit coupling is sizeable
\cite{pand}. In the region around saturation density $\rho_0 \simeq
0.17\,$fm$^3$ it amounts to about half of the "empirical" value $3W_0/4 \simeq
90\,$MeVfm$^5$ deduced in the Skyrme phenomenology of nuclear structure 
\cite{reinhard}. The findings of refs.\cite{pand,pieper} concerning spin-orbit
splittings in light nuclei point of course in the same direction. Finally, the
dashed line in Fig.\,3 shows the isovector spin-orbit strength $G_{\rm
so}(k_f)$ (magnified by a factor 10). In comparison to the isoscalar
spin-orbit strength $F_{\rm so}(k_f)$ it is only a small $5\%$ correction.     

\bigskip
\bild{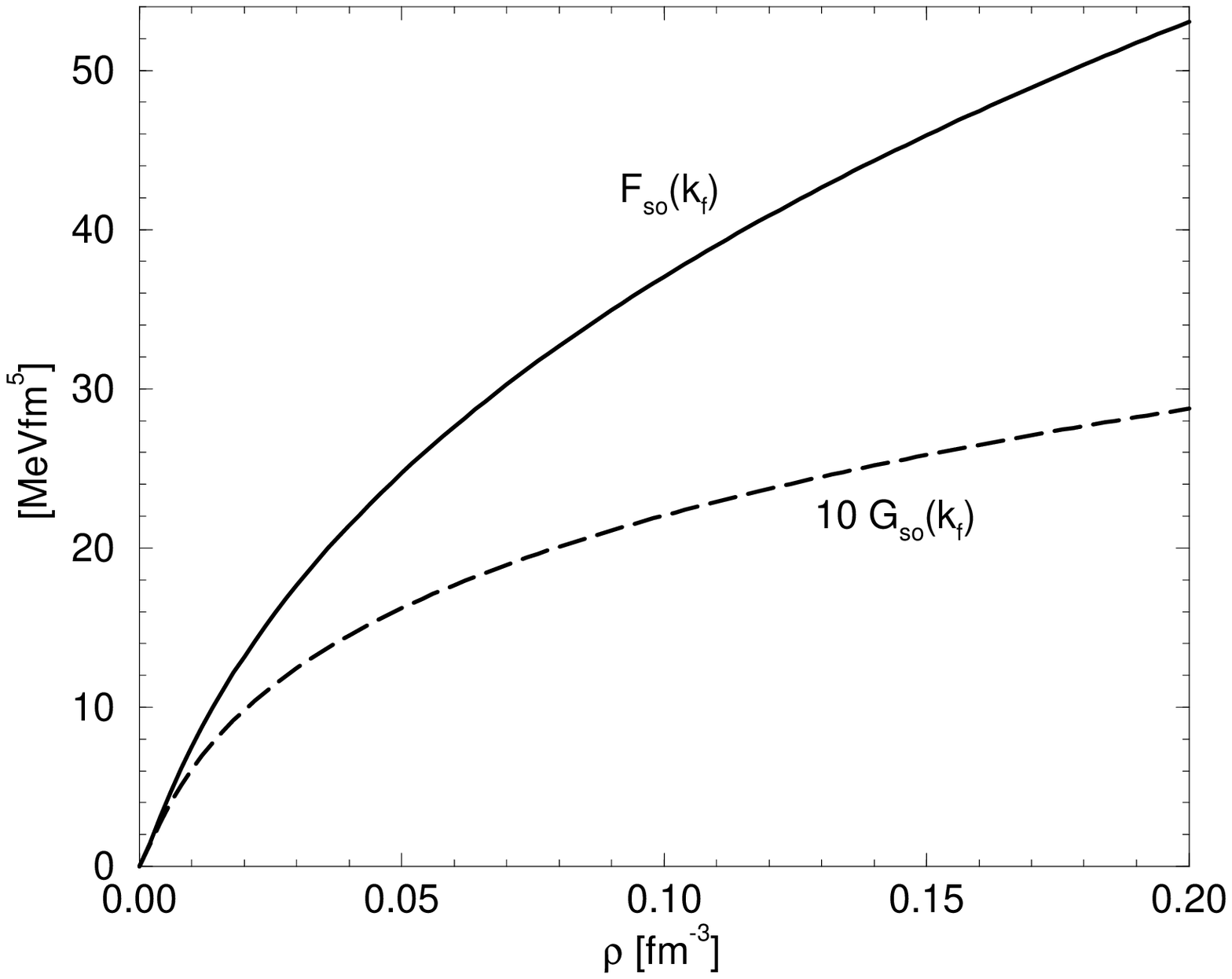}{16}
\vskip -1.3cm
{\it Fig.\,3: The spin-orbit strength generated by the two-pion exchange 
three-nucleon interaction involving virtual $\Delta$-isobar excitation versus
the nucleon density $\rho = 2k_f^3/3\pi^2$. The full curve shows the
isoscalar spin-orbit strength $F_{\rm so}(k_f)$ and the dashed curve shows 
the isovector spin-orbit strength $G_{\rm so}(k_f)$ magnified by a factor 10.}

\bigskip

Next, we turn to a more general derivation of the three-body spin-orbit
coupling generated by two-pion exchange. Chiral symmetry determines the 
$2\pi$-exchange three-nucleon interaction uniquely at leading order 
\cite{friar}. It follows from a tree-diagram involving the chiral $\pi\pi NN
$-contact vertex proportional to the second-order low-energy constants $c_j$:  
\begin{equation} {i \over f_\pi^2} \bigg\{ 2\delta_{ab} \Big[ c_3\, {\vec q}_a
\cdot {\vec q}_b -2c_1\, m_\pi^2 \Big] +c_4\, \epsilon_{abc} \tau_c \,\vec 
\sigma \cdot({\vec q}_a \times {\vec q}_b\,) \bigg\} \,. \end{equation} 
Here, $\vec q_{a,b}$ denote out-going pion-momenta and we have already dropped 
the $c_2$-term proportional to the product of two pion-energies. In the present
application these (off-shell) pion-energies become equal to differences of
nucleon kinetic energies thus producing a higher order relativistic
$1/M^2$-correction. A straightforward evaluation of the (three-loop) Hartree 
diagram in Fig.\,1 with three medium insertions gives the following
contribution to the isoscalar spin-orbit strength: 
\begin{equation} F_{\rm so}(k_f)^{(c_j-\rm Hart)}= {g_A^2 m_\pi \over 4\pi^2
f_\pi^4} \bigg\{{(c_1-c_3)u-2c_3u^3 \over 1+4u^2}+{c_3-c_1\over 4u}\ln(1+4u^2) 
\bigg\} \,, \end{equation} 
where $u=k_f/m_\pi$. The contributions of the s-wave part and the p-wave part 
of the $\pi\pi NN$-contact vertex are distinguished by the two (isoscalar)
low-energy constants $c_1$ and $c_3$. The isovectorial and spin-dependent
$c_4$-vertex makes a non-vanishing contribution only through the Fock
diagram. We find the following total result for the isoscalar spin-orbit
strength from the three-loop Fock diagram  in Fig.\,1: 
\begin{eqnarray} F_{\rm so}(k_f)^{(c_j-\rm Fock)}&=& {g_A^2 m_\pi u^{-3}\over
4\pi^2 (8f_\pi)^4}\bigg\{ -32c_1 \Big[4u-u^{-1} \ln(1+4u^2) \Big]^2 \nonumber 
\\ && -(c_3+c_4) \Big[8u^2-12 +(3u^{-2}+ 4)\ln(1+4u^2) \Big]^2 \bigg\}\,.
\end{eqnarray}  
Its analytical form is remarkably simple. Again there no contribution from the
Hartree diagram in Fig.\,1 to the isovector spin-orbit strength $G_{\rm
so}(k_f)$ while the Fock diagram in Fig.\,1 leads to the combined result: 
\begin{eqnarray} G_{\rm so}(k_f)^{(c_j-\rm Fock)}&=& {g_A^2 m_\pi u^{-3}\over
4\pi^2 (8f_\pi)^4}\bigg\{ -32c_1 \Big[4u-u^{-1} \ln(1+4u^2) \Big]^2 \nonumber 
\\ && +\bigg({c_4\over 3}-c_3\bigg) \Big[8u^2-12 +(3u^{-2}+ 4)\ln(1+4u^2)
\Big]^2 \bigg\}\,. \end{eqnarray}
Here, the isospin structures of the $c_{1,3}$-vertex and the $c_4$-vertex shows
up through relative factors $1$ and $-1/3$. We note aside that the previously 
calculated contributions from explicit $\Delta$-excitation (see eqs.(5,6,7)) 
reappear in the (general) expressions eqs.(11,12,13) through $\Delta$-resonance
contributions to the low-energy constants: $-c_3^{(\Delta)} =2c_4^{(\Delta)} =
g_A^2/2\Delta$. This connection allows one also to trace back the origin of the
factor 7/3 in eq.(7). Finally, we consider the (leading-order) 
Weinberg-Tomozawa $\pi \pi NN$-contact vertex. It generates a three-body 
spin-orbit coupling first in form of a relativistic $1/M$-correction. Because
of the isovector nature of the Weinberg-Tomozawa contact-vertex the Hartree
diagram (in Fig.\,1) can make a contribution only to the isovector spin-orbit
strength: 
\begin{equation} G_{\rm so}(k_f)^{(\rm WT-Hart)}= {g_A^2 m_\pi \over 96\pi^2M 
f_\pi^4} \Big[3\arctan2u-2u -u^{-1}\ln(1+4u^2) \Big] \,. \end{equation}
The Fock diagram (in Fig.\,1) on the other hand generates isoscalar and
isovector spin-orbit strengths in a fixed (relative) ratio: 
\begin{eqnarray} && F_{\rm so}(k_f)^{(\rm WT-Fock)}= -3\, G_{\rm so}(k_f)^{(\rm
WT-Fock)} \nonumber \\ && = {g_A^2 m_\pi u^{-2} \over 4\pi^2 M(4f_\pi)^4}
\Bigg\{ 15u^2 \arctan2u+{3\over 4u}-u-39u^3 \nonumber \\ && -{3+2u^2+10u^4
\over 8u^3}\ln(1+4u^2) +{3+8u^2-64u^4 \over 64u^5} \ln^2(1+4u^2)\nonumber \\ &&
+\int_0^u dx\,x^{-2}\bigg[ 3(u^2+u^4)+ 6(u+u^3)(6x^2-1-u^2) L \nonumber \\ 
&& +\Big[ 3(1+u^2)^3-(15+34u^2+19u^4)x^2 +(33+29u^2)x^4 -13x^6\Big] L^2 
\bigg] \Bigg\} \,, \end{eqnarray}
with the auxiliary function:
\begin{equation} L= {1\over 4x} \ln{1+(u+x)^2 \over 1+(u-x)^2}\,.\end{equation}
The power of the pion mass $m_\pi$ in their prefactors indicates that all
contributions written in eqs.(11-15) are of the same order in the small
momentum expansion. 

\bigskip

\bild{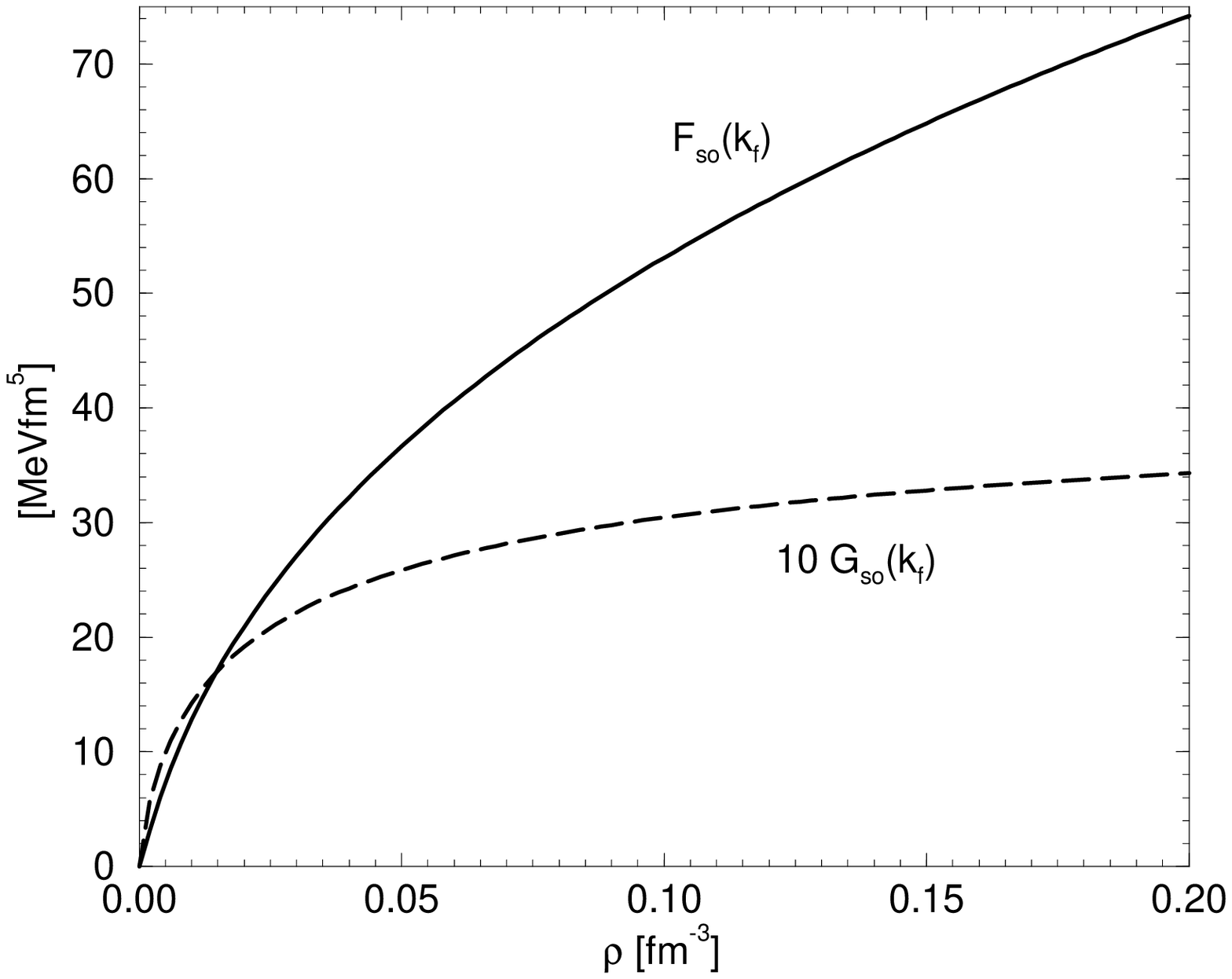}{16}
\vskip-1.3cm
{\it Fig.\,4: The spin-orbit strength generated by the two-pion exchange 
three-nucleon interaction involving the general chiral $\pi\pi NN$-contact 
vertex versus the nucleon density $\rho = 2k_f^3/3\pi^2$. The full curve shows
the isoscalar spin-orbit strength $F_{\rm so}(k_f)$ and the dashed curve shows
the isovector spin-orbit strength $G_{\rm so}(k_f)$ magnified by a factor 10.}

\bigskip

For the numerical evaluation we use the values $c_1 = -0.64\,$GeV$^{-1}$, $c_3
= -3.90\,$GeV$^{-1}$ and $c_4= 2.25\,$GeV$^{-1}$ of the low-energy constants 
which have been determined (at tree-level) in ref.\cite{pipin} from some  
low-energy $\pi N$-data. The full line in Fig.\,4 shows the resulting isoscalar
spin-orbit strength $F_{\rm so}(k_f)$ as a function of the nucleon density 
$\rho = 2k_f^3/3\pi^2$. The contribution of the Hartree diagram involving the 
p-wave contact vertex proportional to $c_3$ is the absolutely dominant one. For
example one finds at nuclear matter saturation density (where $k_f \simeq 2
m_\pi$) the value $F_{\rm so}(2m_\pi)^{(c_3-\rm Hart)}= 65.3\,$MeVfm$^5$. In
comparison to this the s-wave Hartree contribution is very small, $F_{\rm so}(2
m_\pi)^{(c_1- \rm Hart)}= 3.6\,$MeVfm$^5$. Moreover, the even smaller Fock 
contributions have a tendency of cancelling each other. The somewhat smaller 
values of $F_{\rm so}(k_f)$ in Fig.\,3 compared to those in Fig.\,4 originate 
mainly from the fact that the $\Delta$-resonance saturates the low-energy 
constant $c_3$ to about only three quarters in magnitude: $c_3^{(\Delta)} = 
-g_A^2/2 \Delta \simeq -2.9\,$GeV$^{-1}$. The dashed line in Fig.\,4 shows the
isovector spin-orbit strength $G_{\rm so}(k_f)$ (magnified by a factor 10) as
a function of the nucleon density $\rho  =2k_f^3/3\pi^2$. Again, in comparison
to the isoscalar spin-orbit strength $F_{\rm so}(k_f)$ it is only a small 
$5\%$ correction. The largest individual contribution comes here from the Fock 
diagram involving the p-wave $c_3$-contact vertex which for example gives   
at nuclear matter saturation density: $G_{\rm so}(2m_\pi)^{(c_3-\rm Fock)}=
3.2\,$MeVfm$^5$.         
 
In summary we have calculated in this work the spin-orbit coupling generated by
the two-pion exchange three-nucleon interaction. We have made use of the 
density-matrix expansion of Negele and Vautherin \cite{negele}. This method
allows one to compute diagrammatically the nuclear energy density functional 
which contains the spin-orbit coupling term of interest. We have derived simple
analytical expressions for the density-dependent isoscalar and isovector 
spin-orbit strengths $F_{\rm so}(k_f)$ and $G_{\rm so}(k_f)$. First, we have 
considered the two-pion exchange three-nucleon interaction of Fujita and 
Miyazawa \cite{fujita} where one nucleon is excited to a p-wave $\Delta
$-resonance. The corresponding three-loop Hartree and Fock diagrams generate 
spin-orbit couplings which are not relativistic effects but independent of the 
nucleon mass. The Hartree diagram and the isoscalar component $F_{\rm so}(k_f)$
are by far dominant. At nuclear matter saturation density these $\Delta$-driven
three-body mechanisms generate about half of the empirical isoscalar spin-orbit
strength. The calculations of spin-orbit splittings in light nuclei 
\cite{pand,pieper} point of course in the same direction. Secondly, have we 
derived more generally the three-body spin-orbit coupling generated by two-pion
exchange on the basis of the chiral $\pi\pi NN$-contact vertex. In that 
framework we have obtained similar (numerical) results for the density 
dependent isoscalar and isovector spin-orbit strengths $F_{\rm so}(k_f)$ and
$G_{\rm so}(k_f)$. The p-wave part of the chiral $\pi\pi NN$-contact 
interaction (proportional to the low-energy constant $c_3$) and the Hartree 
diagram give rise to the absolutely dominant contribution. 

On the other hand it has been shown recently in ref.\cite{efun} that iterated 
one-pion exchange generates an isoscalar spin-orbit strength $F_{\rm so}(k_f)$ 
that is sizeable but of the wrong negative sign. Combining those results
\cite{efun} with the present ones, one may conclude that the net nuclear
spin-orbit coupling generated by (multi) pion-exchange is rather small, at
least for densities around nuclear matter saturation density. Lorentz scalar
and vector mean-fields with their in-medium behavior governed by QCD sum rules
could therefore be the appropriate dynamical framework for building up the
strong (isoscalar) nuclear spin-orbit interaction. Indeed such a proposal has
recently been successfully applied in ref.\cite{finelli} to nuclear structure
calculations.


\begin{thebibliography}{99}
\bibitem{walecka} B.D. Serot and J.D. Walecka, {\it Int. J. Mod. Phys.} {\bf 
E6}, 515 (1997); and references therein.\vs 
\bibitem{ringreview} P. Ring, {\it Prog. Part. Nucl. Phys.} {\bf 37}, 193
(1996); and references therein.\vs
\bibitem{fujita} J. Fujita and H. Miyazawa, {\it Prog. Theor. Phys.} {\bf 17},
360 (1957); {\bf 17}, 366 (1957).\vs
\bibitem{pand} S.C. Pieper and V.R. Pandharipande, {\it Phys. Rev. Lett.} {\bf
70}, 2541 (1993).\vs
\bibitem{akmal} A. Akmal, V.R. Pandharipande and D.G. Ravenhall, {\it
Phys. Rev.} {\bf C58}, 1804 (1998).\vs
\bibitem{pieper} S.C. Pieper, V.R. Pandharipande, R.B. Wiringa, and J. Carlson,
{\it Phys. Rev.} {\bf C64}, 014001 (2001); and references therein.\vs
\bibitem{friar} J.L. Friar, D. H\"uber, and U. van Kolck, {\it Phys. Rev.} {\bf
C59}, 53 (1999).\vs 
\bibitem{negele} J.W. Negele and D. Vautherin, {\it Phys. Rev.} {\bf C5}, 1472
(1972).\vs
\bibitem{reinhard} M. Bender, P.-H. Heenen and P.-G. Reinhard,  {\it Rev. Mod. 
Phys.} {\bf  75}, 121 (2003); and references therein.\vs
\bibitem{efun} N. Kaiser, S. Fritsch and W. Weise, \textit{Nucl. Phys.} 
\textbf{A724}, 47 (2003); and references therein.\vs
\bibitem{delta} N. Kaiser, S. Gerstend\"orfer and W. Weise, \textit{Nucl. 
Phys.} \textbf{A637}, 395 (1998).\vs
\bibitem{pipin} V. Bernard, N. Kaiser and Ulf-G. Mei{\ss}ner, \textit{Nucl. 
Phys.} \textbf{B457}, 147 (1995).\vs 
\bibitem{finelli} P. Finelli, N. Kaiser, D. Vretenar and W. Weise,
\textit{Eur. Phys. J.} \textbf{A17}, 573 (2003); nucl-th/0307069.\vs 
\end{thebibliography}
\end{document}